\documentclass[12pt]{article}
\usepackage{graphicx}
\usepackage{amsmath}
\usepackage[geometry]{ifsym} 
\usepackage{bbm}
\usepackage{amsfonts}
\usepackage{enumerate}
\usepackage{subfig}
\usepackage{natbib}
\usepackage[nolists]{endfloat}

\usepackage{algorithmic}
\usepackage{algorithm}
\newcommand{\bs}{\boldsymbol}

\begin{document}

\title{Signal extraction and breakpoint identification for \\ 
array CGH data using robust state space model}
\author{Bin Zhu, Jeremy M.G. Taylor and Peter X.-K. Song \\
Department of Biostatistics \\ University of Michigan}

\date{}
\maketitle

\centerline{\textbf{Abstract}}
Array comparative genomic hybridization(CGH) is a high resolution technique to assess DNA copy number variation. Identifying breakpoints where copy number changes will enhance the understanding of the pathogenesis of human diseases, such as cancers. However, the biological variation and experimental errors contained in    array CGH data may lead to false positive identification of breakpoints.
We propose a robust state space model for array CGH data analysis. The model consists of two equations: an observation equation and a state equation, in which both the measurement error and evolution error are specified to follow $t$-distributions with small degrees of freedom. The completely unspecified CGH profiles are estimated by a Markov Chain Monte Carlo(MCMC) algorithm. Breakpoints and outliers are identified by a novel backward selection procedure based on posterior draws of the CGH profiles. Compared to three other popular methods, our method demonstrates several desired features, including false positive rate control, robustness against outliers, and superior power of breakpoint detection. All these properties are illustrated using simulated and real datasets.



\newpage

\section{Introduction}
Almost all types of cancer share one common characteristic, genetic instability, including DNA copy number variation(CNV).  
During cancer progression some genes will lose one of the two copies or are
completely deleted, while others may gain one copy, or become amplified up to hundreds of copies. These chromosomal alterations can lead to abnormal cell proliferation, DNA repair, senescence and apoptotic mechanisms and can provide a selective advantage for cells and result in cancer. Identification of CNV not only enhances the understanding of oncogenesis but also facilitates the treatment of cancer. For example, Trastuzumab is a monoclonal antibody interfering with ERBB2 receptor and is used for the treatment of breast cancers with amplified, and multiple copies of the ERBB2 gene \citep{vogel2002efficacy}.

Array comparative genomic hybridization (CGH) is a technique that is used to detect differences in DNA copy number \citep{solinastoldo1997mbc, pinkel1998hra}. The isolated DNA from tumor and the normal tissue from each patient are labeled with different fluorescent dyes and then cohybridized to the microarray. The $\log_2$ fluorescent intensity ratios are measured at different chromosomal positions to define each CGH profile. This CGH profile is supposed to be proportional to the copy number ratio for tumor and normal cells across the chromosome.  See \citet{pinkel2005acg} and \citet{davies2005act} for detail reviews.
Array CGH data exhibit three challenging characteristics. First, the data displays abrupt changes at the positions where DNA copy number is possibly
altered. Second, the data usually contain biological variations and experimental errors, which hinder the accurate identification of breakpoints where copy number changes. Biological variations refer to heterogeneity of copy number within tumor cells and  experimental errors include contamination of the tumor cells with normal cells, measurement errors and errors caused by processing tissue samples. Third, the data are spatially dependent. That is, neighboring genes are more likely to share the same copy number than remote ones. The primary aim of array CGH data analysis is to estimate the CGH profiles and to identify breakpoints from available noisy observations.

A number of statistical methods have been proposed for array CGH data analysis. Most of the methods postulate that the observed $\log_2$ intensity ratio $Y(t_j)$ is governed by the following model,
\begin{equation}
\label{eq:obs.n}
Y(t_j) = \mu(t_j) + \varepsilon(t_j),\;\;\;j=1,2,\dots,J
\end{equation}
where signal $\mu(t_j)$ is the true $\log_2$ intensity ratio at $j$th probe, $\varepsilon(t_j)$ is noise and $t_j$ denotes the physical position of $j$th probe  on a chromosome. Different assumptions and interpretations of $\mu(t_j)$ and $\varepsilon(t_j)$ lead to various estimation approaches, which may be categorized into three types. The first type is based on the segmentation method. It assumes that the CGH profile ${\mu}(t)$,  is piecewise constant, i.e. 
$\mu(t_j) = \sum_{m=1}^{M} \mu_m I[t_j \in \mathcal{T}_m]$, where $\mathcal{T}_m$ is segment $m$ with mean $\mu_m$ and $I(\cdot)$ is the indicator function. Also $\varepsilon(t_j)$ follows independent and identically distributed (i.i.d.) $\mathcal{N}(0, \sigma^{2}_{\varepsilon})$. 
To detect breakpoints that enable us to classify chromosome into blocks, \citet{olshen2004cbs} and \citet{venkatraman2007fcb} proposed the method of circular binary segmentation(CBS); \citet{hupe2004aac} developed the adaptive weighted smoothing procedure; and \citet{erdman2008fbc} implemented a Bayesian change point model.
The second type is the method of hidden Markov models (HMM), which restricts $\bs{\mu}(t)$ to take a finite number of values and uses a Markov chain to model   probabilities: $\Pr(\mu(t_{j+1}) = \mu_{m^\prime} \mid \mu(t_j) = \mu_m )$, $\mu_m,\mu_{m^\prime} \in \mathcal{U}$, with state space $\mathcal{U}=\{\mu_m; m = 1, 2, \dots,M \}$. Note that $M$ is a prespecified number of states. The HMM method was first applied to array CGH data analysis by \citet{fridlyand2004hmm}. \citet{shah2006} modified the HMM method to achieve robustness against outliers. A  continuous-index HMM was developed by \citet{stj2007cih}. \citet{guha2008bhm} derived a Bayesian approach to the HMM with objective decision rules. A segmental maximum posteriori approach(SMAP) by \citet{andrew2008smp} has incorporated both
genomic distance and overlap between clones into the HMM. 
Finally, the third type is built upon penalization methods, which essentially relax the piecewise constant assumption by imposing a roughness penalty on CGH profile ${\mu}$(t). In a penalization method, we consider minimizing an objective function of the form 
$
Q=Q_{gf}+Q_{sp},
$
where the first term $Q_{gf}$ measures the goodness of fit for profile $\mu(t)$ to the observed process $Y(t)$ at observed probes $t_j'$s, and the second term $Q_{sp}$ regularizes the smoothness of $\mu(t)$. Various forms of $Q_{gf}$ and $Q_{sp}$ have been proposed in the literature, including quantile smoothing \citep{eilers2005qsa}, LASSO \citep{huang2005ddc}, fused quantile regression\citep{li2007aac} and fused LASSO (FLASSO) by \citet{Tibshirani2008}. 

Besides the three types of methods, there are other approaches; for example, clustering algorithm \citep{wang2005mcg,liu2006dbc}, wavelet transform \citep{hsu2005dab} and ridge regression \citep{vandewiel2009swa}, among many others. 
Comprehensive comparisons among some of aforementioned methods were given by \citet{lai2005ca} and \citet{wk2005}. Some of the methods only estimate the profiles but do not directly call the breakpoints. Further, most methods do not control the false positive rate for breakpoint identification, and their performances are significantly effected by the experimental errors, such as outliers.     
In this paper, we propose a new method based on robust state space models for array CGH data to estimate the CGH profile and to identify breakpoints under controlled false positive rates. In addition, this new method has a number of desirable properties: (1) it is robust against outliers; (2) it incorporates physical distance between probes into CNV identification; (3) it enables us to quantify estimation uncertainties of signals via posterior credible intervals;  (4) all the parameters are estimated as part of the MCMC algorithm and thus are highly data-adaptive; (5) the computational efficiency of the MCMC algorithm for profile estimation is proportional to the number of probes, which helps the computation speed for high-throughput array CGH data analysis.

The rest of the paper is organized as follows. In Section 2, we first present the robust state space model, then describe an MCMC algorithm to draw samples of both profiles and parameters, and also outline a novel procedure of calling the breakpoints and outliers using MCMC samples. In Section 3, the proposed model and  method are applied to both simulated and real datasets for illustration, where we compare our new method to three popular existing methods.  We finally give conclusions and discussion in Section 4. 

\section{Methods}

\subsection{Model}
For the ease of exposition, we denote $Y_j = Y(t_j)$, $\mu_j=\mu(t_j)$ and $\varepsilon_j = \varepsilon(t_j)$. The proposed robust state space model(RSSM) comprises two equations: an observation equation and a state equation. The observation equation is given in equation (\ref{eq:obs.n}), where error $\varepsilon_j$ is assumed to be i.i.d and  follow $t$-distribution, $\mathcal{T}_{\upsilon_{\varepsilon}}$, with degree of freedom(d.f.) $\upsilon_{\varepsilon}$.
Note that $t$-distribution is a scale-mixture of normal distribution and gamma distribution. Thus, we rewrite $\varepsilon_j \sim \mathcal{N}(0, \sigma^{2}_{\varepsilon,j})$ a normal distribution with mean $0$ and variance $\sigma^{2}_{\varepsilon,j}$,  and let 
$\sigma^{-2}_{\varepsilon,j} = \lambda_{\varepsilon,j} \tau_{\varepsilon}$ and
$\lambda_{\varepsilon,j} \sim \mathcal{G}(\upsilon_{\varepsilon}/2,\upsilon_{\varepsilon}/2)$ a gamma distribution with shape parameter $\upsilon_{\varepsilon}/2$ and rate parameter $\upsilon_{\varepsilon}/2$. The priors are specified as 
$\upsilon_{\varepsilon} \sim \mathcal{G}(10^{-3},10^{-3})I(2,10)$ and 
$\tau_{\varepsilon}\sim \mathcal{G}(10^{-3},10^{-3})$ throughout the paper.

We regard the signal $\mu(t_j)$ as a continuous quantity which measures the $\log_2$ of average copy number of heterogeneous tumor cells versus homogeneous normal cells. The state equation is:
\begin{equation}
\label{eq:state.n}
\mu_{j+1} = \mu_{j} + \xi_{j},
\end{equation}
where the evolution error or signal difference $\xi_{j}$ follows an i.i.d $t$-distribution with d.f $\upsilon_{\xi}$. Similar to the specification of $\varepsilon_j$, we let $\xi_{j} \sim \mathcal{N}(0, \sigma^2_{\xi,j} \delta_j)$, $\delta_j = t_{j+1} - t_{j}$, $\sigma^{-2}_{\xi,j} = \lambda_{\xi,j} \tau_{\xi}$ and  
$\lambda_{\xi,j} \sim \mathcal{G}(\upsilon_{\xi}/2,\upsilon_{\xi}/2)$, with the priors
$\upsilon_{\xi} \sim \mathcal{G}(10^{-3},10^{-3})I(0.01,2)$ and 
$\tau_{\xi}\sim \mathcal{G}(10^{-3},10^{-3})$. As a result,
$\varepsilon_{j} \sim \mathcal{T}_{\upsilon_{\varepsilon}}(0,\tau_{\varepsilon}^{-1})$ and 
$\xi_{j} \sim \mathcal{T}_{\upsilon_{\xi}}(0,\delta_j \tau_{\xi}^{-1})$ marginally. Unlike other robust state space models \citep{west1984outlier,fahrmeir1999penalized}, our model incorporates the physical distance $\delta_j$ between two probes to address the feature that the farther two probes are apart, the larger the signal difference $\xi_j$ is likely to be. Note that degree of freedom $\upsilon_{\xi}$ is limited  below 2. In this way, we hope that the distribution $\mathcal{T}_{\upsilon_{\xi}}$ can accommodate extremely large values of signal difference probably caused by breakpoints. A similar strategy was suggest by \citet{kitagawa1987non}, where differences of signals are modeled by a distribution in the Pearson system with no finite second moments. As shown in his paper, the Pearson system distribution facilitates the detection of mean structure changes.


\subsection{Signal extraction by MCMC}
With the model formulation given in Section 2.1, we now outline an MCMC algorithm to sample from the posterior distribution for signals $\bs{\mu}=[\mu_1,\mu_2,\dots,\mu_J]^\top$, parameters $\bs{\phi_o}=[\lambda_{\varepsilon, j}, \upsilon_{\varepsilon},\tau_{\varepsilon}]$ and $\bs{\phi_s}=[\lambda_{\xi,j},\upsilon_{\xi},\tau_{\xi}]$ for $j=1,2,\dots,J$, given the data $\bs{Y}=[Y_1,Y_2,\dots,Y_J]$.

\begin{itemize}
\item Given $\bs{Y}$, $\bs{\phi_o}$ and $\bs{\phi_s}$, update the $\bs{\mu}$ by the simulation smoother \citep{db2002}, a multi-state Gibbs sampler which very efficiently draws samples from the posterior distribution of signals  $\bs{\mu}$.

\item Given $\bs{Y}$ and $\bs{\mu}$, update $\bs{\phi_o}$ according to the following steps:

$[\lambda_{\varepsilon,j} \mid \cdotp] \sim 
\mathcal{G}(\frac{\upsilon_{\varepsilon}}{2} + \frac{1}{2},
\frac{\upsilon_{\varepsilon}}{2} + \frac{(Y_{j} - \mu_{j})^2\tau_{\varepsilon}}{2})$;

$[\upsilon_{\varepsilon} \mid \cdotp] = 
\prod_{j=1}^{J} 
\mathcal{G}(\lambda_{\varepsilon j} \mid \frac{\upsilon_{\varepsilon}}{2}, \frac{\upsilon_{\varepsilon}}{2})
\mathcal{G}(\upsilon_{\varepsilon} \mid 10^{-3}, 10^{-3})I(2,10)
$, by Adaptive Metropolis Rejection Sampling (ARMS; \citealp{gilks1995});

$\tau_{\varepsilon} \sim
\mathcal{G}(\frac{J}{2}+ 10^{-3}, \sum_{j=1}^{J}\frac{(Y_j - \mu_{j})^2 \lambda_{\varepsilon j}}{2}+ 10^{-3}).
$
\item Given $\bs{\mu}$, update $\bs{\phi_s}$ through the following steps:

$[\lambda_{\xi,j} \mid \cdotp] \sim 
\mathcal{G}(\frac{\upsilon_{\xi}}{2} + \frac{1}{2},
\frac{\upsilon_{\xi}}{2} + \frac{(\mu_{j+1} - \mu_{j})^2\tau_{\xi}}{2 \delta_j})$;

$[\upsilon_{\xi} \mid \cdotp] = 
\prod_{j=1}^{J} 
\mathcal{G}(\lambda_{\xi j} \mid \frac{\upsilon_{\xi}}{2}, \frac{\upsilon_{\xi}}{2})
\mathcal{G}(\upsilon_{\xi} \mid 10^{-3}, 10^{-3})I(0.01,2)
$, by the ARMS;

$\tau_{\xi} \sim
\mathcal{G}(\frac{J}{2}+ 10^{-3}, \sum_{j=1}^{J}\frac{(\mu_{j+1} - \mu_{j})^2 \lambda_{\xi j}}{2 \delta_j}+ 10^{-3})
$.

\end{itemize} 
According to the definition of errors $\varepsilon_j=Y_j-\mu_j$ and $\xi_j=\mu_{j+1}-\mu_j$, 
we obtain the posterior draws of the errors $\bs{\varepsilon}=[\varepsilon_1,\varepsilon_2,\dots,\varepsilon_J]^\top$ and the signal differences $\bs{\xi}=[\xi_2,\xi_3,\dots,\xi_J]^\top$. Samples of $\bs{\varepsilon}$ and  $\bs{\xi}$ are essential to identify outliers and breakpoints through a novel backward selection procedure detailed in Section 2.3 below.   
 
\subsection{Breakpoints and outliers calling}
Breakpoints are called by our backward selection procedure outlined in Algorithm \ref{alg1} given in the Supplementary Material. The input to the algorithm is the posterior draws of signal differences $\bs{\xi}$, in an $m \times n$ matrix, with $m$ denoting the number of draws and $n$ equal to the number of probes minus one, as well as an input of a threshold $q_\alpha$. The specification of $q_\alpha$ is discussed in detail below. At line \ref{line:6} in Algorithm \ref{alg1}, we calculate  $\tilde{P}_j$, which is an estimate of the posterior probability $P[| \xi_j 
| > | \xi_{-j}| \mid \bs{Y}]$. This is the probability of the absolute value of signal difference at position $j$ is larger than those at any other positions, given the data. The quantity $P[| \xi_j 
| > | \xi_{-j}| \mid \bs{Y}]$ represents the area under the ROC curve or AUC \citep[Ch.4]{pepe2004statistical}. It is known that AUC measures the separation between the posterior distribution of $|\xi_j|$ and that of the remaining $|\xi_{-j}|$, namely all $|\xi_{i}|$ with $i \neq j$. Under the null hypothesis that probe $j$ is not a breakpoint, we expect $\tilde{P}_j$ to be near $0.5$. The decision of rejection of the null hypothesis will be based on the comparison of $\tilde{P}_j$ with the threshold $q_\alpha$. In the first iteration of procedure, several $\tilde{P}_j'$s may be larger than $q_\alpha$; we take the largest one and call it a breakpoint. This called position will be excluded from the subsequent iterations. we repeat his calling procedure for the remaining $\xi_j$ until none of the remaining $\tilde{P}_j$ is above the threshold $q_\alpha$ or all $n-1$ breakpoints are selected. The output of the algorithm is a list of identified breakpoints. Likewise, we utilize this backward selection procedure to call outliers, based on the posterior draws of errors $\bs{\varepsilon}$.

In the above backward selection procedure for the calling of breakpoints, the 
$q_\alpha$ is determined such that under the null hypothesis that probe $i$ is not a breakpoint, it will be chosen with probability $\alpha$(i.e.  $\alpha$ is false positive rate). When a normal reference array is available, we can measure $\log_2$ intensity ratio of normal versus normal tissue. Fitting the proposed state space model to the normal reference array, we obtain the posterior draws of signal differences $\bs{\xi}^o$, where we can obtain $\tilde{P}_j^{o'}$s according to Algorithm \ref{alg1}. These $\tilde{P}_j^o$, $j=1,2,\dots,J^o$, can be regarded as a random sample from a distribution under the null hypothesis. Then, the $q_\alpha$ is obtained as the  $(1-\alpha)$ quantile of all $\tilde{P}_j^{o\prime}$s. This quantile $q_\alpha$ implies that under the null hypothesis, the rate of false positive is $\alpha$.
In some real experiments,  normal reference arrays however may not be available. In this case, we can generate a pseudo normal reference array $\bs{Y}^o=[Y^o_1,Y^o_2,\dots,Y^o_{J^o}]$ by sampling with replacement from the data $\bs{Y}$. In this case, if some $Y_j'$s in the aberration region are sampled, they will be dispersed and scattered randomly within the set $\bs{Y}^o$. Thus, they appear most likely as outliers rather than a contiguous pattern of changes. Since the proposed state space model is robust against outliers, the $q_\alpha$ under the null hypothesis can be reasonably determined.
Given the pseudo normal array, the steps to obtain the $q_\alpha$ are the same as those given in the scenario of the normal reference array being available. 
          
\section{Applications}

\subsection{Simulation study}
We first evaluate our proposed method and compare it with three other popular methods, FLASSO, CBS and SMAP, using well known artificial chromosomes simulated by \citet{lai2005ca}. Lai \textit{et al.}'s data consist of  100 chromosomes, each with length 100.
In the center of each chromosome is added an aberration of copy number gain, which has one of the four different width (5,10,20 and 40).
The signal-to-noise ratio(SNR) is 1,  and the noise follows a normal distribution with standard deviation 0.25.

\begin{figure}[b]
 \centering
\includegraphics[width=0.85\textwidth,angle=270]{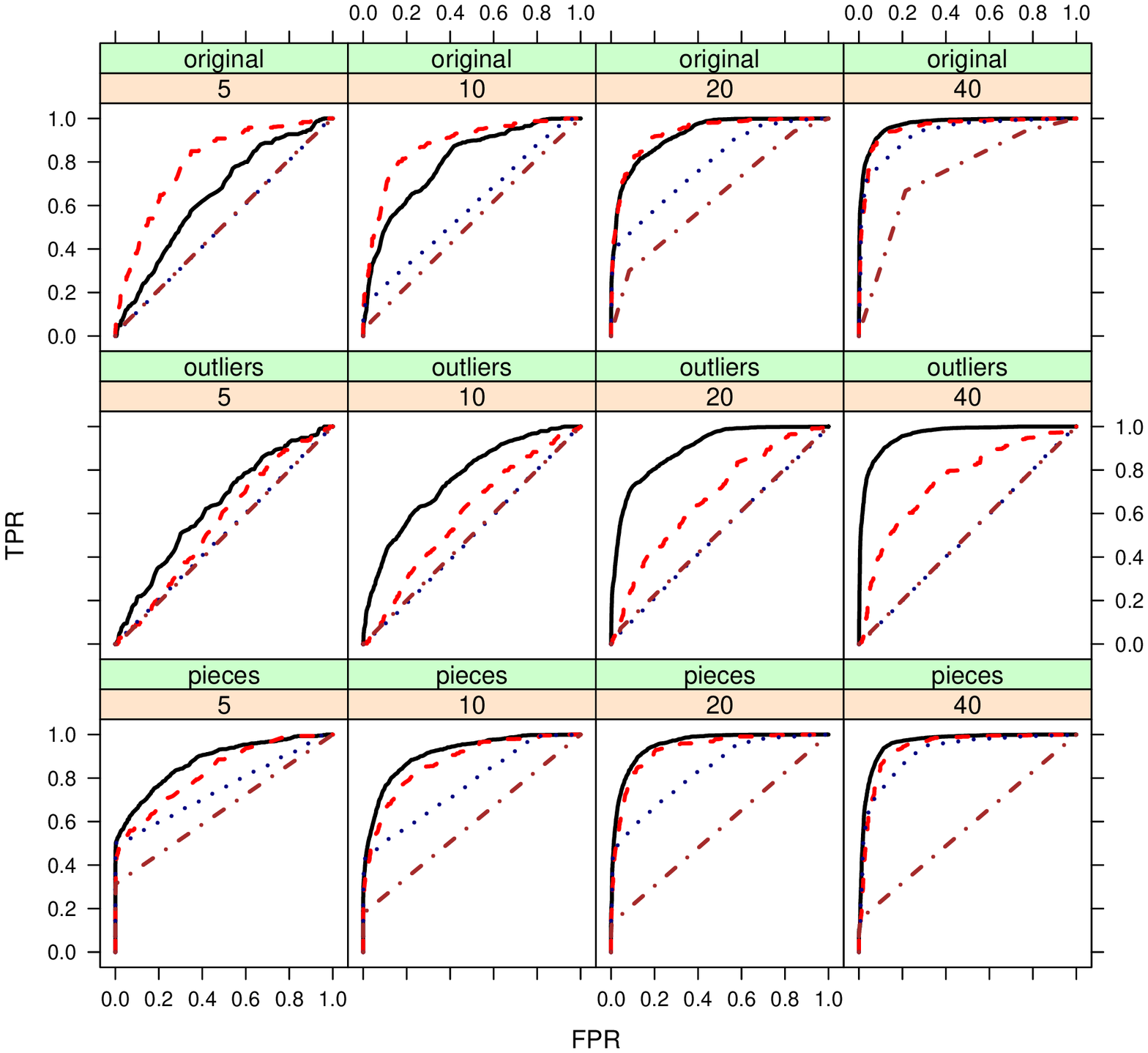}
   \caption{ROC curves of four methods at SNR 1. \textemdash \; Our model,$---$ FLASSO, $-\; \cdotp -$ CBS, $\cdotp \cdotp \cdotp$ SMAP.\label{fig:LaiROC1}}
\end{figure}

We use the Receiver Operating Characteristic(ROC) curve to compare the performance of the four methods in each width case. To obtain ROC curves, we compare the estimated signal $\hat{\mu}_j$ at each location with a cutoff varying from the minimum to the maximum of $\bs{Y}$, and regard the location $i$ where $\hat{\mu}_i$ is above the cutoff as the detected aberration region. The true positive rate(TPR) is defined as the proportion of the true aberration region detected as an aberration region, while the false positive rate(FPR) is defined as the proportion of the normal region declared as an aberration region. The TPRs and FPRs are plotted as ROC curves in Figure \ref{fig:LaiROC1}. For the Lai \textit{et al.}'s data, the plots at the first row in Figure \ref{fig:LaiROC1} indicate that our approach performs clearly better than CBS and SMAP methods, in terms of higher TPR and lower FPR, not as well as FLASSO for the narrow regions but comparably to FLASSO for the wide aberrations(20 and 40).

The simulated data in \citet{lai2005ca} is idealized, and does not contain any of the complex features that occur in real data.  Outliers are commonly seen in real datasets for various reasons, including single probe amplification/deletion or experimental errors. To investigate the effect of outliers, in Lai \textit{et al.}'s simulated dataset, we add five percent of outliers in each chromosome at randomly selected positions with magnitudes uniformly distributed over interval $(3,6)$. The ROC curves given at the second row in Figure \ref{fig:LaiROC1} clearly show the advantage of the proposed method. Comparing to the corresponding cases in the first row, the ROC curves of FLASSO, CBS and SMAP are considerably closer to the diagonal line, demonstrating a significant loss of prediction power for the detection of CNVs. In contrast, the ROC curve of the proposed approach is affected very little, indicating clearly that our method is robust to outliers.

Another feature of the real data is the possibility of more than one region of aberration with different magnitudes. To evaluate the performance of the methods, we explore cases when two aberration regions are present in the simulated chromosome simultaneously. For each Lai \textit{et al.}'s simulated chromosome, a randomly selected normal region of width five is replaced by an aberration block with SNR 4. Based on the ROC curves plotted in the third row of Figure \ref{fig:LaiROC1}, the proposed approach outperforms the three other approaches. 

\begin{figure}[b]
 \centering
\includegraphics[width=0.85\textwidth,angle=270]{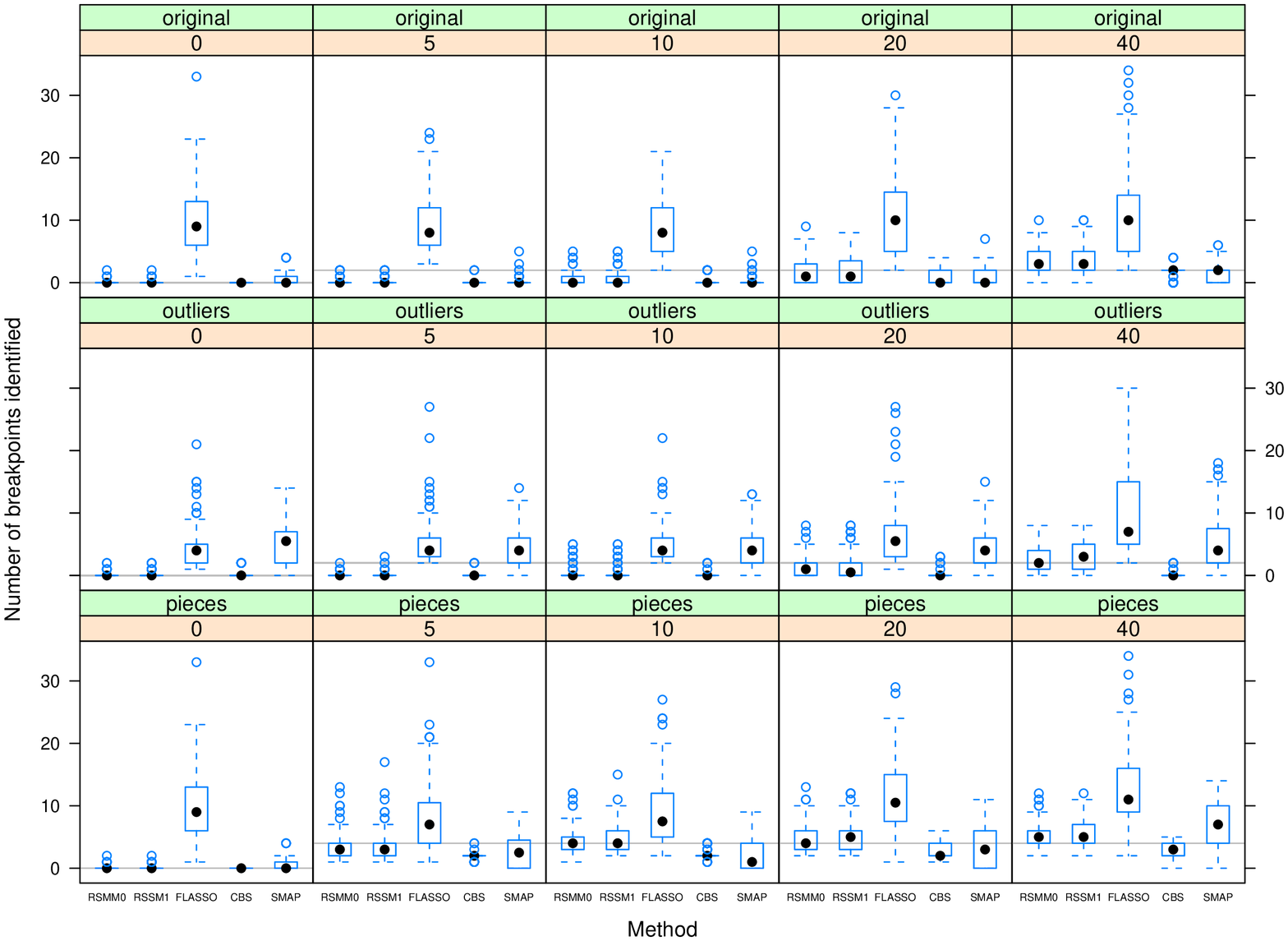}
   \caption{Breakpoints identification using simulated and pseudo normal reference arrays. The horizontal reference lines indicate the true number of breakpoints. The simulated normal reference array is labeled by  RSSM0, while RSSM1 utilizes pseudo normal reference arrays based on resampling the observed data. The panel on bottom left is the replicate of the one on top left. \label{fig:HR.rep}}
\end{figure}

An important task in array CGH analysis is to correctly identify breakpoints.
We investigate the number of breakpoints identified by the four methods for each chromosome in the above simulated data. In addition, we simulated normal chromosomes without any aberration regions. For these we generate 100 normal chromosomes, each with 100 probes simulated from $\mathcal{N}(0,0.25^2)$. In addition, another 100 chromosomes are generated by adding outliers to the 100 normal chromosomes, in the same way described above. For FLASSO, CBS and SMAP, a breakpoint is defined as a position $j$, if the difference is non-zero,  that is, $\Delta \hat{\mu}_j =\hat{\mu}_{j+1} - \hat{\mu}_j \neq 0$. For the proposed method, a breakpoint is called by the backward selection procedure as described in Algorithm \ref{alg1}. To determine the $q_\alpha$, we simulate a normal reference array with each probe as $\mathcal{N}(0,0.25^2)$ with length $J^o=1000$ and generate the pseudo normal reference arrays with length $J^o=1000$ through sampling with replacement from the artificial chromosomes.  The false positive rate $\alpha$ is set at 0.001, which means that for every 1000 probes in the normal reference array, one probe is expected to be falsely called as a breakpoint. Figure \ref{fig:HR.rep} shows the side-by-side boxplots of the number of breakpoints identified by each of four method respectively, where the $q_\alpha$ is determined with simulated normal reference arrays and pseudo normal reference arrays, respectively, for RSSM0 and RSSM1 corresponding to the first two boxplots in each panel. From a comparison of  these boxplots, it is clear that the number of breakpoints is over-estimated substantially by FLASSO in all the three scenarios although the magnitude of the signal difference at some of these breakpoints may by quite small. The true number of breakpoints,  on average,  is more likely to be correctly achieved by the proposed method, in  scenario of two pieces of aberration regions or in the cases where the aberration widths are as wide as 20 and 40. For the normal chromosomes with or without outliers, both CBS and our method correctly conclude that there are no breakpoints, while FLASSO identifies a few number of false breakpoints. Note that our method identifies a total of 6 and 13 breakpoints for 10,000 probes in 100 normal chromosomes by using, respectively, simulated and pseudo normal reference arrays.  These number of false discoveries numbers are close to the expected number 10, given the false positive rate 0.001. We also notice that the numbers of breakpoints identified in the simulated and pseudo normal reference arrays are very close to each other, which validates the utility of pseudo normal reference arrays when the normal reference arrays are not available.          

\subsection{Glioblastoma Multiforme(GBM) data}
\begin{figure}[b]
  \centering
  \subfloat[Chromosome 7 in GBM 29]
           {\includegraphics[width=0.40\textwidth,angle=0]{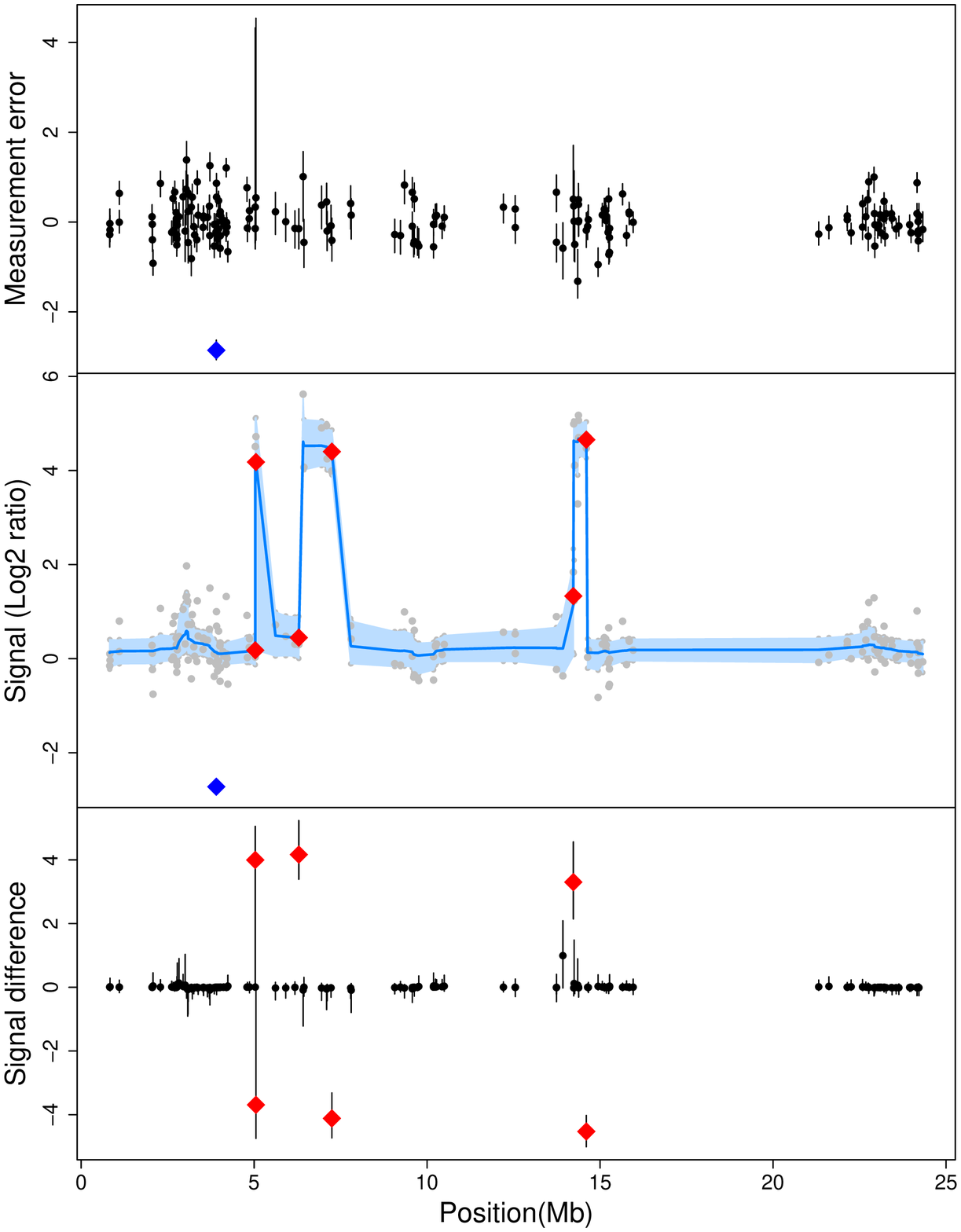}}
  \subfloat[Chromosome 13 in GBM31]
           {\includegraphics[width=0.40\textwidth,angle=0]{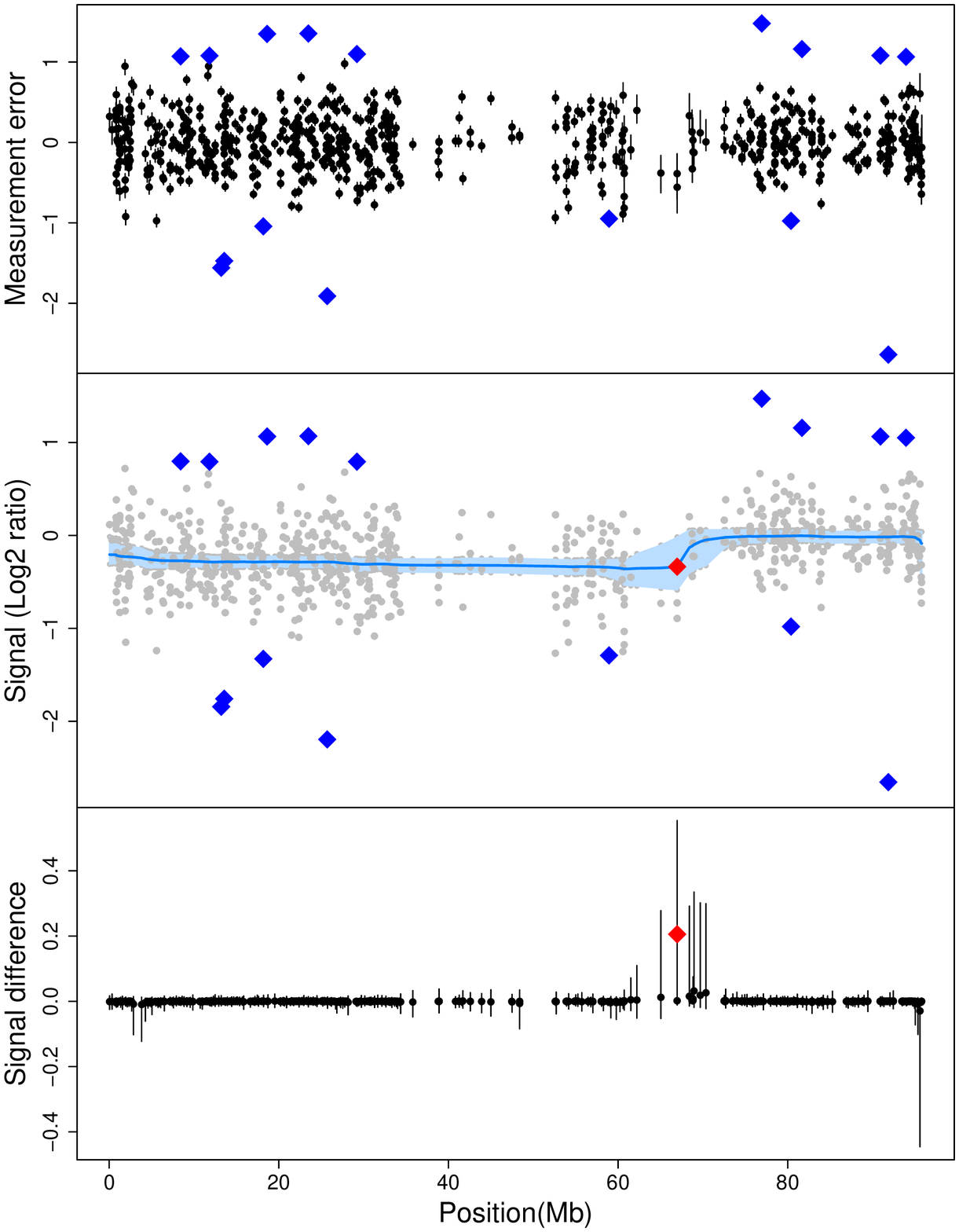}}                
  
    \caption{GBM panel plots for the posterior distributions of measurement error, signal, and signal difference by state space model. In the top and bottom panels, the $\bullet$ denotes the posterior mean and $\mid$ stands for the $95\%$ credible intervals. In the middle panel, gray $\bullet$ is the data point and \textemdash is posterior mean and $95\%$ credible intervals are the shaded areas. \FilledSmallDiamondshape denotes the selected outliers and breakpoints.    
\label{fig:GBM}}
\end{figure}

\begin{figure}[b]
 \centering
\includegraphics[width=0.85\textwidth,angle=270]{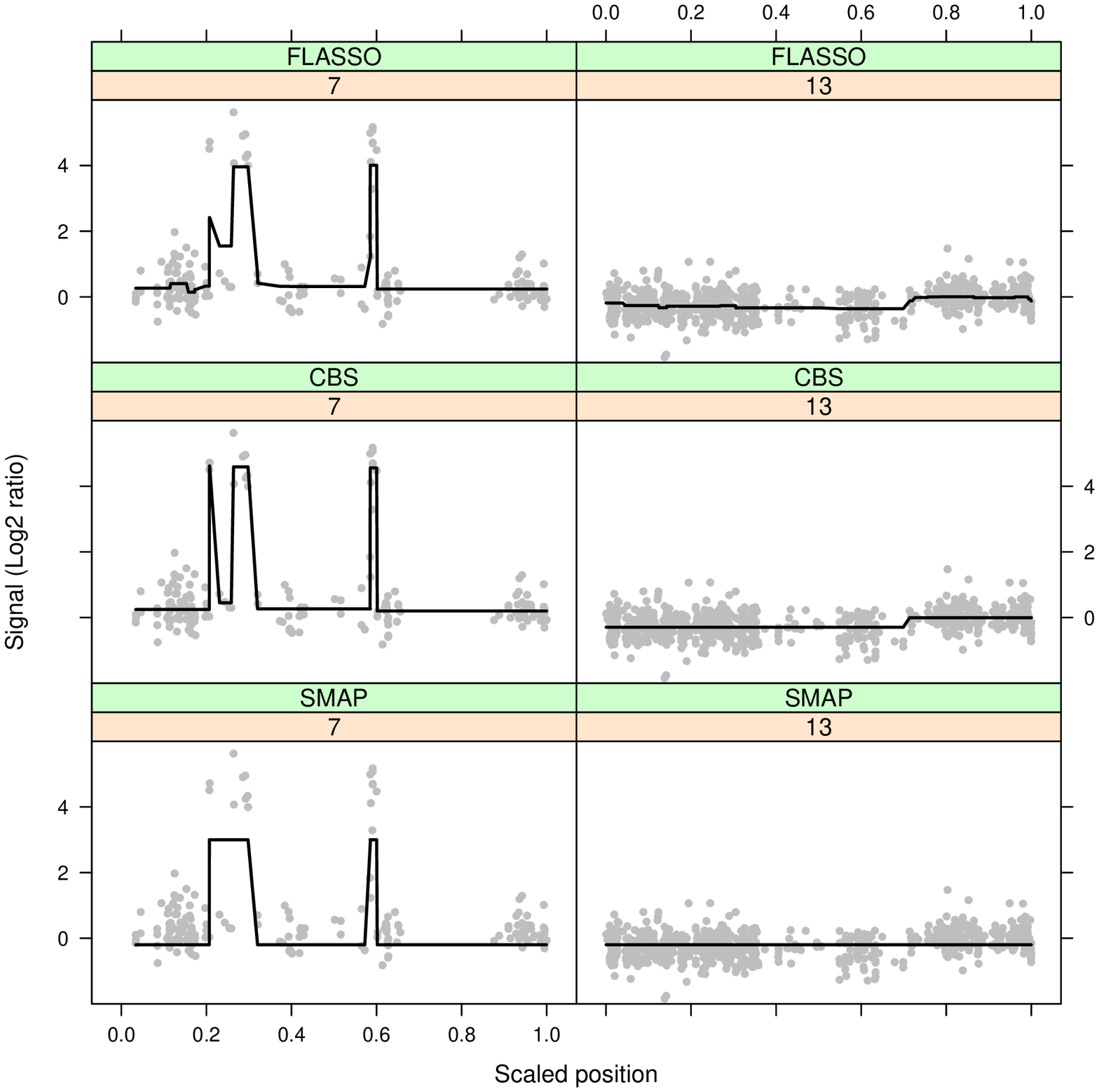}
   \caption{Panel plots of signal(\textemdash) estimated for GBM data by FLASSO, CBS and SMAP, where gray $\bullet$ denotes the data point.\label{fig:GBM.others}}
\end{figure}

GBM data by \citet {bredel2005high} include 26 samples representing primary GBMs, the most malignant type of brain tumor. In sample GBM31, a large region of loss is demonstrated on chromosome 13, which is also observed by \citet{koschny2002comparative} in a meta-analysis of 509 cases.  Besides losses, the GBM data also contain a number of amplifications, one of which is shown on  chromosome 7 in sample GBM29. \citet{lai2005ca} compared the performance of various methods based on these two chromosomes 13 and 7 with challenging features. They represent wider, low level region of loss,  and narrower, high level region of amplification, respectively. To assess our proposed method, we re-analyze these two chromosomes using our method. The analysis is based on 1000 MCMC draws from a single chain of $75,000$ iterations with $25,000$ burn-in period and every 50th being recorded.  As shown in Figure \ref{fig:GBM} , our method successfully detects both the loss region and amplification region as well as some outliers. Both breakpoints and outliers are called using the proposed backward selection procedure. The threshold for breakpoints is obtained through the pseudo normal reference arrays with $q^\xi_{0.001}=0.911$ for chomosome 7 and $q^\xi_{0.001}=0.882$ for chomosome 13.  The threshold for outliers is chosen as $q^\varepsilon=0.98$. The panels in Figure \ref{fig:GBM} also illustrate posterior means and $95\%$ credible intervals for signal $\mu_j$, error $\varepsilon_j$ and signal differences $\xi_j$ across the chromosomes. At a given position, the wider interval indicates higher uncertainty. Note that $95\%$ credible intervals of signal difference illustrate the corresponding posterior distributions. The further the credible interval departs from the others along with the narrower width, the stronger it indicates the corresponding position is a breakpoint. We also analyze the GBM data using the  methods of FLASSO, CBS and SMAP. As Figure \ref{fig:GBM.others} shown, all three methods can identify the two aberration regions, except SMAP method that fails to detect any aberration region for chromosome 13. 

Table \ref{Tab:02} lists the number of breakpoints identified by each of the four methods. Our method and CBS reach the same numbers on both chromosomes, which are much less than the those found by FLASSO.     

\begin{table}[!t]
\caption{The number of breakpoints identified in GBM and Breast Tumor data \label{Tab:02}}
{\begin{tabular}{llllllll}\hline
\hline
&\multicolumn{2}{c}{GBM data}
&
&\multicolumn{4}{c}{Breast Tumor data}\\
\cline{2-3}
\cline{5-8}
&CH7&CH13&&CH8&CH11&CH17&CH20\\
\hline
RSSM(ours)&6&1&&9&5&3&5\\
FLASSO&15&15&&30&19&29&12\\
CBS&6&1&&3&6&0&2\\
SMAP&4&0&&6&11&8&9\\
\hline
\end{tabular}
}
\end{table} 
  
\subsection{Breast tumor data}
\begin{figure}[b]
  \centering
  \subfloat[Chromosome 8]
           {\includegraphics[width=0.40\textwidth,angle=0]{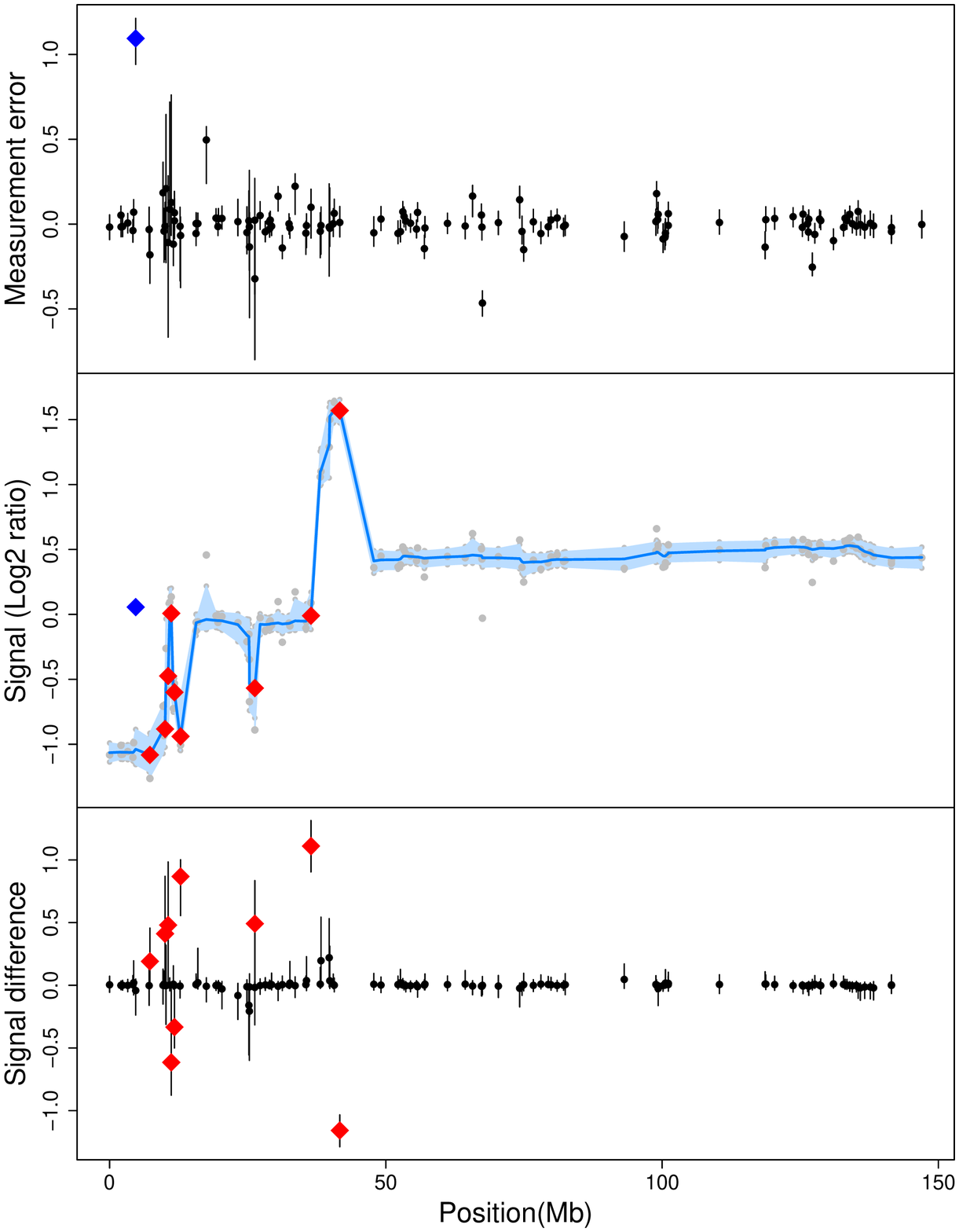}}                
  \subfloat[Chromosome 11]
           {\includegraphics[width=0.40\textwidth,angle=0]{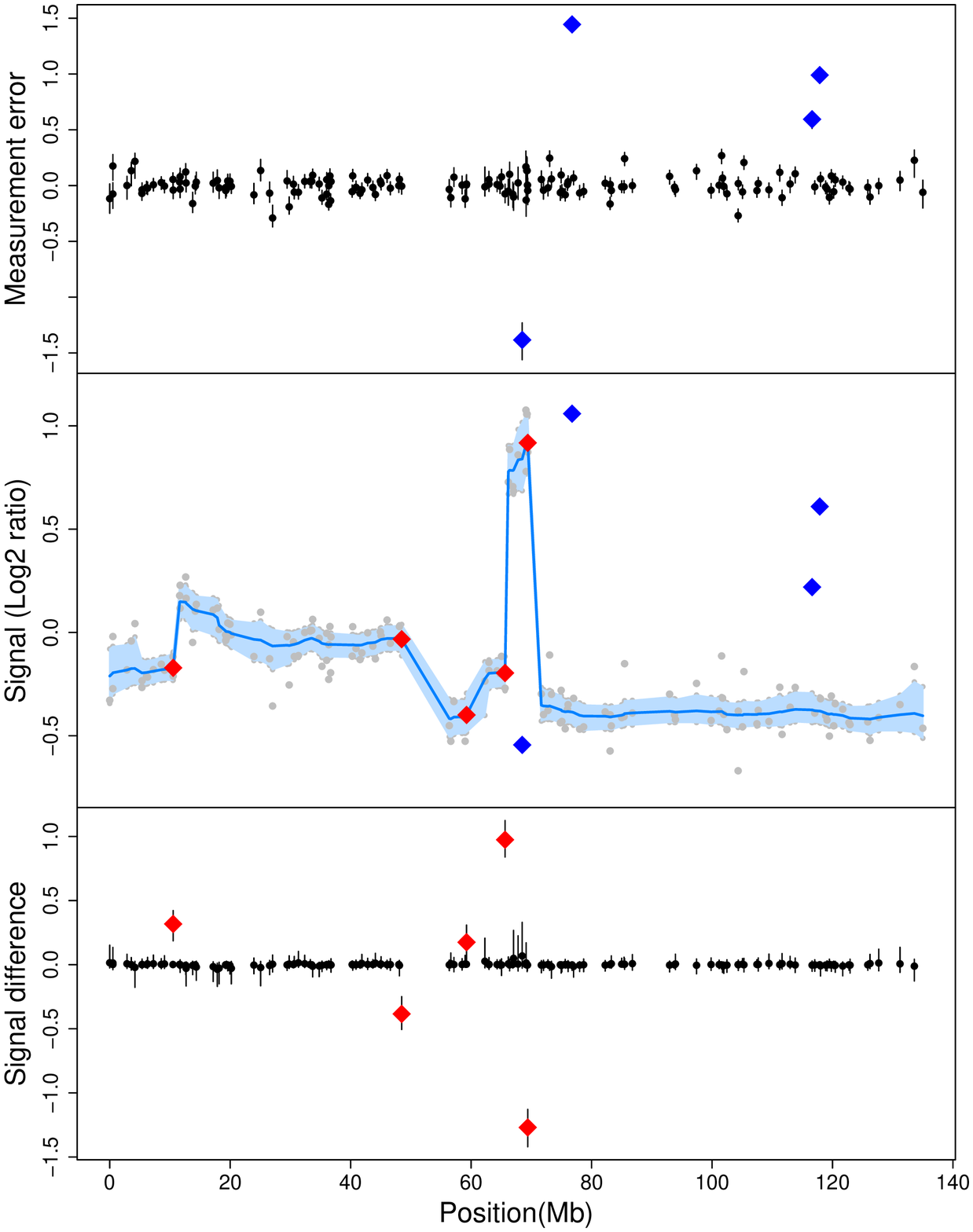}}\\
  \subfloat[Chromosome 17]
           {\includegraphics[width=0.40\textwidth,angle=0]{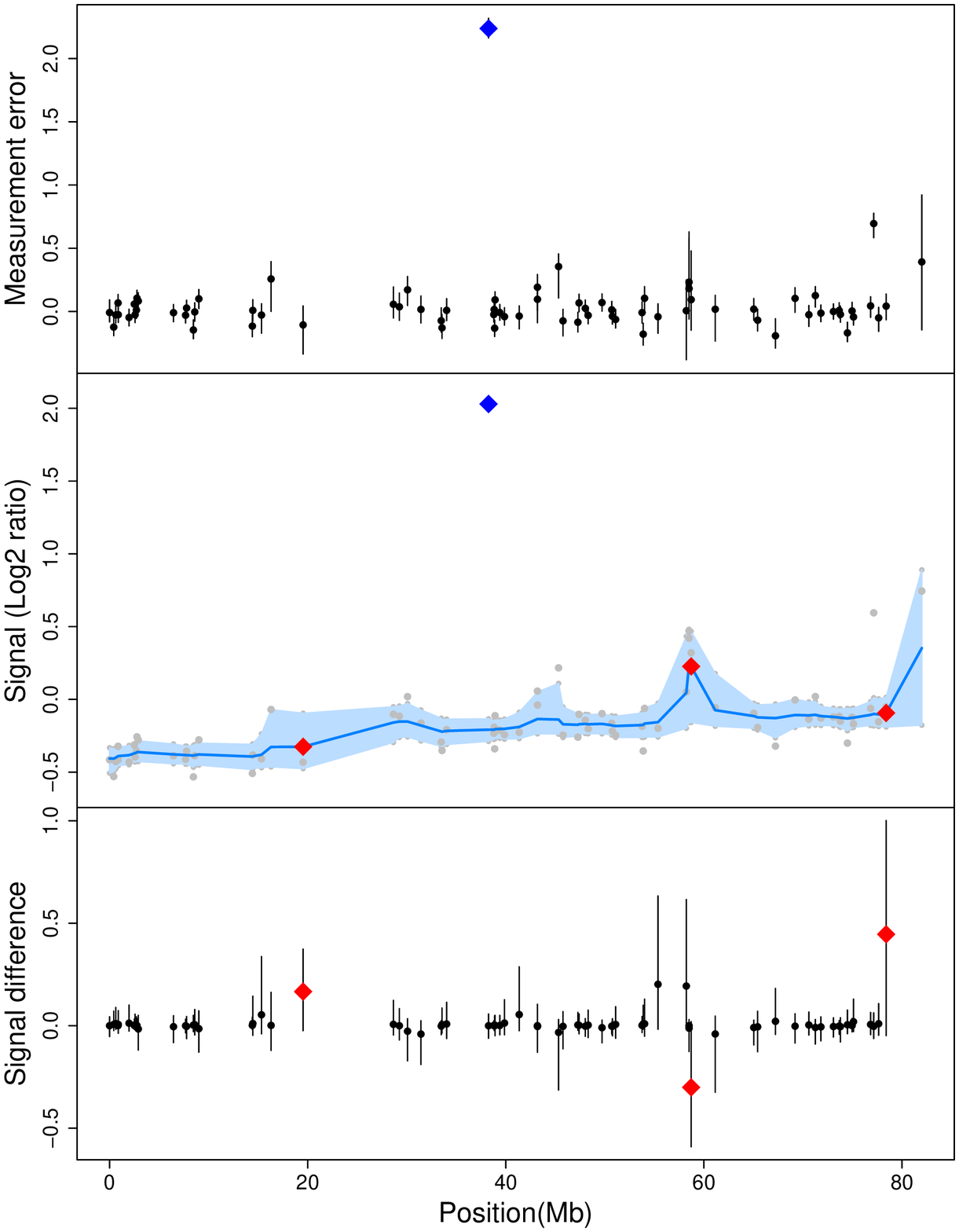}}
  \subfloat[Chromosome 20]
           {\includegraphics[width=0.40\textwidth,angle=0]{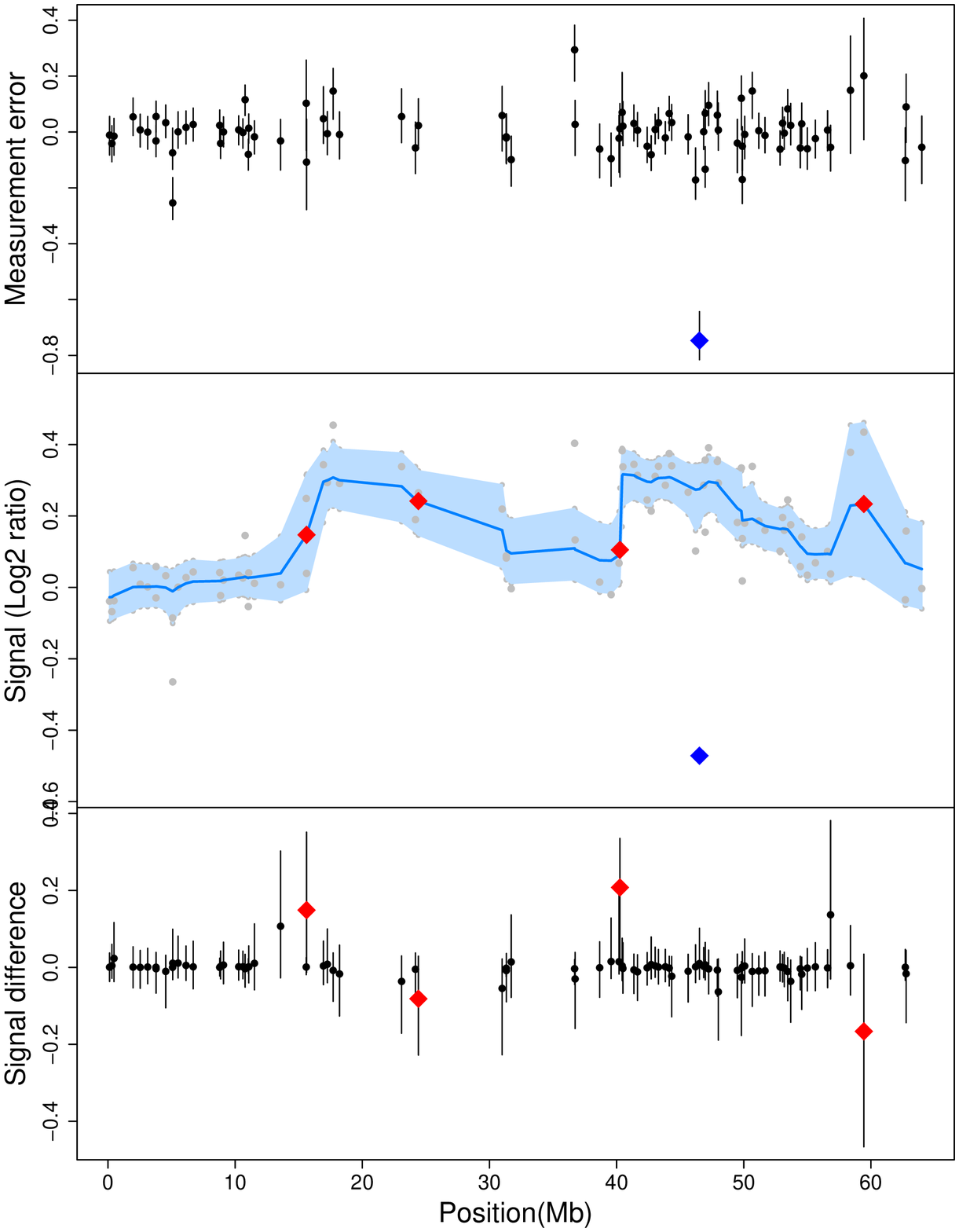}}
    \caption{Breast tumor panel plots for the posterior distributions of measurement error, signal, and signal difference by state space model. In the top and bottom panels, the $\bullet$ denotes the posterior mean and $\mid$ stands for the $95\%$ credible intervals. In the middle panel, gray $\bullet$ is the data point and \textemdash is posterior mean and $95\%$ credible intervals are the shaded areas. \FilledSmallDiamondshape denotes the selected outliers and breakpoints.  
\label{fig:BC.ssm}}
\end{figure}

\citet{fridlyand2006breast} considered array CGH data from across 2464 genomic clones in 62 sporadic ductal invasive breast tumors and 5 BRCA1 mutant tumors. We apply our method as well as other three methods to analyze four chromosomes(8,11,17 and 20) of tumor "S1539", in which there are a number of low level gains and losses as well as high level amplifications. The results of our method are  based on 1000 MCMC draws from a single chain of $75,000$ iterations with $25,000$ burn-in period and every 50th being recorded. The backward selection procedure has been applied to identify a number of breakpoints and outliers/amplifications. The $q^{\varepsilon}$ is specified as 0.98, and $q_{0.001}^{\xi}$ is determined using the pseudo normal reference arrays, resulting in values of $0.795,0.789,0.808$ and $0.807$ for chromosome 8,11,17 and 20 respectively.
Figure \ref{fig:BC.ssm} displays the posterior means and $95\%$ credible intervals of signal $\mu_j$, error $\varepsilon_j$ and signal differences $\xi_j$ across the chromosomes, as well as a number of called outliers and breakpoints.
These breakpoints define the edges of aberration regions which include several well-known oncogenes, that play key roles in the pathogenesis of breast tumor. The detected regions cover gene FGFR I between 36.4Mb and 39.7Mb on chromosome 8, gene CCND I between 68.5Mb and 77.0Mb on chromosome 11, and gene ZNF217 between 44.4Mb and 62.7Mb on chromosome 20. Gene ERBB2 between 34.1Mb and 38.7Mb on chromosome 17 is a well known gene that can be amplified in breast cancer. There are very few probes close to ERBB2, and the method detected a probe as an outlier in this region.  

\begin{figure}[b]
 \centering
\includegraphics[width=0.85\textwidth,angle=270]{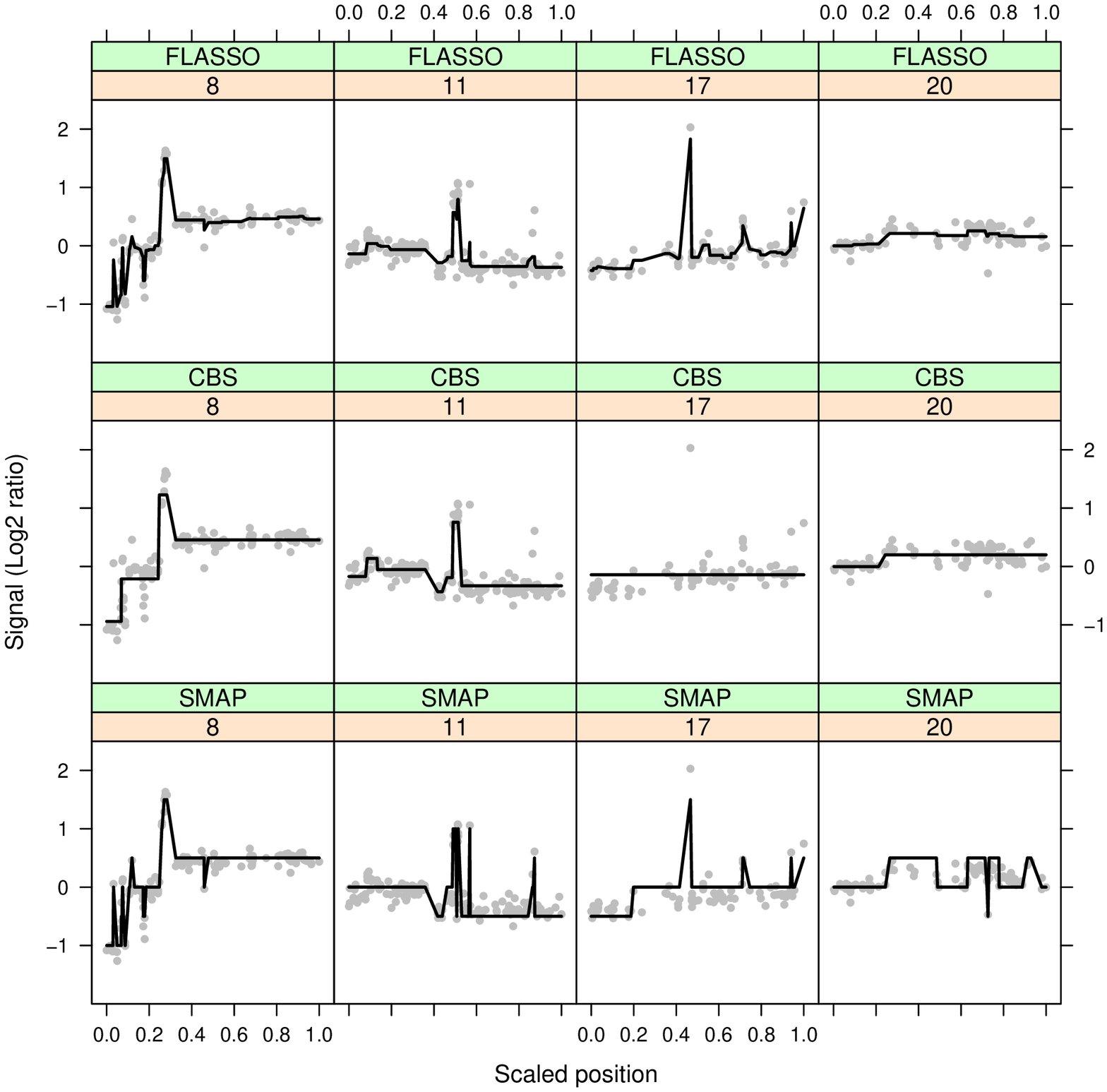}
   \caption{Panel plots of signal(\textemdash) estimated for breast tumor data by FLASSO, CBS and SMAP, where gray $\bullet$ denotes the data point.\label{fig:BC.others}}
\end{figure}

We also analyze the same breast tumor data by FLASSO, CBS and SMAP methods. The results are shown in Figure \ref{fig:BC.others}. We can see that the SMAP method appears to be very sensitive to outliers(e.g. in chromosome 11) and local features(e.g. in chromosome 20), which has obscured the estimate of the global trend. The CBS method failed to capture the single probe amplification in the chromosome 17 and the weak gain in chromosome 20. The FLASSO method is also sensitive to outliers, e.g. at the beginning of chromosome 8 and in the middle of chromosome 11.  The number of breakpoints identified by each of the methods is summarized in Table \ref{Tab:02}. FLASSO identifies a large number of breakpoints, our method identifies slightly more breakpoints than CBS and slightly fewer than SMAP.

\section{Discussion}
In this paper, we have proposed a powerful new method based on a robust state space model to detect CNVs from array CGH data. A key feature of the proposed method is the use of heavy tail $t$-distributions, which facilitates the robustness in the calling of breakpoints and outliers. Through an MCMC algorithm, our approach presents an appealing method for CGH profile estimation and detection of breakpoints.
Our method is based on a probability model that gives not only point estimation, but also uncertainty intervals for the signal, signal difference and measurement error magnitudes, as illustrated in Figure \ref{fig:GBM} and \ref{fig:BC.ssm}. Such displays are very useful for visualizing the data and the degree of confidence in any conclusion. 
We developed a novel backward selection procedure to effectively utilize the MCMC samples in the identification of breakpoints and outliers/amplifications. Importantly, we control the false positive rate of feature detection at a prespecified level by using real or pseudo normal reference arrays. As illustrated by both simulated and real datasets, our approach has demonstrated superior detection power for aberration regions and breakpoints, and outperforms other existing methods in most of cases, especially for noisy data with outliers. 


\bibliographystyle{bioinformatics}
\bibliography{document}

\begin{thebibliography}{}

\bibitem[Andersson {\em et~al.}, 2008]{andrew2008smp}
Andersson, R., Bruder, C., Piotrowski, A., Menzel, U., Nord, H., Sandgren, J.,
  Hvidsten, T., de~Stahl, D., Dumanski, J.  \& Komorowski, J. (2008{\em{}}) {A
  segmental maximum a posteriori approach to genome-wide copy number
  profiling}.
\newblock {\em Bioinformatics, } {\bf 24} (6), 751.

\bibitem[Bredel {\em et~al.}, 2005]{bredel2005high}
Bredel, M., Bredel, C., Juric, D., Harsh, G., Vogel, H., Recht, L.  \& Sikic,
  B. (2005{\em{}}) {High-resolution genome-wide mapping of genetic alterations
  in human glial brain tumors}.
\newblock {\em Cancer Research, } {\bf 65} (10), 4088.

\bibitem[Davies {\em et~al.}, 2005]{davies2005act}
Davies, J., Wilson, I.  \& Lam, W. (2005{\em{}}) {Array CGH technologies and
  their applications to cancer genomes}.
\newblock {\em Chromosome Research, } {\bf 13} (3), 237--248.

\bibitem[Durbin \& Koopman, 2002]{db2002}
Durbin, J. \& Koopman, S. (2002{\em{}}) {A simple and efficient simulation
  smoother for state space time series analysis}.
\newblock {\em Biometrika, } {\bf 89} (3), 603--616.

\bibitem[Eilers \& De~Menezes, 2005]{eilers2005qsa}
Eilers, P. \& De~Menezes, R. (2005{\em{}}) {Quantile smoothing of array CGH
  data}.
\newblock {\em Bioinformatics, } {\bf 21} (7), 1146--1153.

\bibitem[Erdman \& Emerson, 2008]{erdman2008fbc}
Erdman, C. \& Emerson, J. (2008{\em{}}) {A fast Bayesian change point analysis
  for the segmentation of microarray data}.
\newblock {\em Bioinformatics, } {\bf 24} (19), 2143.

\bibitem[Fahrmeir \& K\"{u}nstler, 1999]{fahrmeir1999penalized}
Fahrmeir, L. \& K\"{u}nstler, R. (1999{\em{}}) {Penalized likelihood smoothing
  in robust state space models}.
\newblock {\em Metrika, } {\bf 49} (3), 173--191.

\bibitem[Fridlyand {\em et~al.}, 2004]{fridlyand2004hmm}
Fridlyand, J., Snijders, A., Pinkel, D., Albertson, D.  \& Jain, A.
  (2004{\em{}}) {Hidden Markov models approach to the analysis of array CGH
  data}.
\newblock {\em Journal of Multivariate Analysis, } {\bf 90} (1), 132--153.

\bibitem[Fridlyand {\em et~al.}, 2006]{fridlyand2006breast}
Fridlyand, J., Snijders, A., Ylstra, B., Li, H., Olshen, A., Segraves, R.,
  Dairkee, S., Tokuyasu, T., Ljung, B., Jain, A.  {\em et~al.} (2006{\em{}})
  {Breast tumor copy number aberration phenotypes and genomic instability}.
\newblock {\em BMC Cancer, } {\bf 6} (1), 96.

\bibitem[Gilks {\em et~al.}, 1995]{gilks1995}
Gilks, W., Best, N.  \& Tan, K. (1995{\em{}}) {Adaptive rejection Metropolis
  sampling within Gibbs sampling}.
\newblock {\em Applied Statistics, } {\bf 44} (4), 455--472.

\bibitem[Guha {\em et~al.}, 2008]{guha2008bhm}
Guha, S., Li, Y.  \& Neuberg, D. (2008{\em{}}) {Bayesian hidden Markov modeling
  of array CGH data}.
\newblock {\em Journal of the American Statistical Association, } {\bf 103}
  (482), 485--497.

\bibitem[Hsu {\em et~al.}, 2005]{hsu2005dab}
Hsu, L., Self, S., Grove, D., Randolph, T., Wang, K., Delrow, J., Loo, L.  \&
  Porter, P. (2005{\em{}}) {Denoising array-based comparative genomic
  hybridization data using wavelets}.
\newblock {\em Biostatistics, } {\bf 6} (2), 211--226.

\bibitem[Huang {\em et~al.}, 2005]{huang2005ddc}
Huang, T., Wu, B., Lizardi, P.  \& Zhao, H. (2005{\em{}}) {Detection of DNA
  copy number alterations using penalized least squares regression}.
\newblock {\em Bioinformatics, } {\bf 21} (20), 3811--3817.

\bibitem[Hupe {\em et~al.}, 2004]{hupe2004aac}
Hupe, P., Stransky, N., Thiery, J., Radvanyi, F.  \& Barillot, E. (2004{\em{}})
  {Analysis of array CGH data: from signal ratio to gain and loss of DNA
  regions}.
\newblock {\em Bioinformatics, } {\bf 20} (18), 3413.

\bibitem[Kitagawa, 1987]{kitagawa1987non}
Kitagawa, G. (1987{\em{}}) {Non-Gaussian state-space modeling of nonstationary
  time series}.
\newblock {\em Journal of the American Statistical Association, } {\bf 82},
  1032--1041.

\bibitem[Koschny {\em et~al.}, 2002]{koschny2002comparative}
Koschny, R., Koschny, T., Froster, U., Krupp, W.  \& Zuber, M. (2002{\em{}})
  {Comparative genomic hybridization in glioma a meta-analysis of 509 cases}.
\newblock {\em Cancer genetics and Cytogenetics, } {\bf 135} (2), 147--159.

\bibitem[Lai {\em et~al.}, 2005]{lai2005ca}
Lai, W., Johnson, M., Kucherlapati, R.  \& Park, P. (2005{\em{}}) {Comparative
  analysis of algorithms for identifying amplifications and deletions in array
  CGH data}.
\newblock {\em Bioinformatics, } {\bf 21} (19), 3763--3770.

\bibitem[Li \& Zhu, 2007]{li2007aac}
Li, Y. \& Zhu, J. (2007{\em{}}) {Analysis of array CGH data for cancer studies
  using fused quantile regression}.
\newblock {\em Bioinformatics, } {\bf 23} (18), 2470.

\bibitem[Liu {\em et~al.}, 2006]{liu2006dbc}
Liu, J., Mohammed, J., Carter, J., Ranka, S., Kahveci, T.  \& Baudis, M.
  (2006{\em{}}) {Distance-based clustering of CGH data}.
\newblock {\em Bioinformatics, } {\bf 22} (16), 1971.

\bibitem[Olshen {\em et~al.}, 2004]{olshen2004cbs}
Olshen, A., Venkatraman, E., Lucito, R.  \& Wigler, M. (2004{\em{}}) {Circular
  binary segmentation for the analysis of array-based DNA copy number data}.
\newblock {\em Biostatistics, } {\bf 5} (4), 557--572.

\bibitem[Pepe, 2004]{pepe2004statistical}
Pepe, M. (2004{\em{}}) {\em {The statistical evaluation of medical tests for
  classification and prediction}}.
\newblock Oxford University Press, USA.

\bibitem[Pinkel \& Albertson, 2005]{pinkel2005acg}
Pinkel, D. \& Albertson, D. (2005{\em{}}) {Array comparative genomic
  hybridization and its applications in cancer}.
\newblock {\em Nature Genetics, } {\bf 37} (6), S11--S17.

\bibitem[Pinkel {\em et~al.}, 1998]{pinkel1998hra}
Pinkel, D., Segraves, R., Sudar, D., Clark, S., Poole, I., Kowbel, D., Collins,
  C., Kuo, W., Chen, C., Zhai, Y.  {\em et~al.} (1998{\em{}}) {High resolution
  analysis of DNA copy number variation using comparative genomic hybridization
  to microarrays}.
\newblock {\em Nature Genetics, } {\bf 20} (2), 207--211.

\bibitem[Shah {\em et~al.}, 2006]{shah2006}
Shah, S., Xuan, X., DeLeeuw, R., Khojasteh, M., Lam, W., Ng, R.  \& Murphy, K.
  (2006{\em{}}) {Integrating copy number polymorphisms into array CGH analysis
  using a robust HMM}.
\newblock {\em Bioinformatics, } {\bf 22} (14).

\bibitem[Solinas-Toldo {\em et~al.}, 1997]{solinastoldo1997mbc}
Solinas-Toldo, S., Lampel, S., Stilgenbauer, S., Nickolenko, J., Benner, A.,
  Dohner, H., Cremer, T.  \& Lichter, P. (1997{\em{}}) {Matrix-based
  comparative genomic hybridization: biochips to screen for genomic
  imbalances}.
\newblock {\em Genes, Chromosomes and Cancer, } {\bf 20} (4).

\bibitem[Stjernqvist {\em et~al.}, 2007]{stj2007cih}
Stjernqvist, S., Ryden, T., Skold, M.  \& Staaf, J. (2007{\em{}})
  {Continuous-index hidden Markov modelling of array CGH copy number data}.
\newblock {\em Bioinformatics, } {\bf 23} (8), 1006.

\bibitem[Tibshirani \& Wang, 2008]{Tibshirani2008}
Tibshirani, R. \& Wang, P. (2008{\em{}}) {Spatial smoothing and hot spot
  detection for CGH data using the fused lasso}.
\newblock {\em Biostatistics, } {\bf 9} (1), 18--29.

\bibitem[van~de Wiel {\em et~al.}, 2009]{vandewiel2009swa}
van~de Wiel, M., Brosens, R., Eilers, P., Kumps, C., Meijer, G., Menten, B.,
  Sistermans, E., Speleman, F., Timmerman, M.  \& Ylstra, B. (2009{\em{}})
  {Smoothing waves in array CGH tumor profiles}.
\newblock {\em Bioinformatics, } {\bf 25} (9), 1099--1104.

\bibitem[Venkatraman \& Olshen, 2007]{venkatraman2007fcb}
Venkatraman, E. \& Olshen, A. (2007{\em{}}) {A faster circular binary
  segmentation algorithm for the analysis of array CGH data}.
\newblock {\em Bioinformatics, } {\bf 23} (6), 657.

\bibitem[Vogel {\em et~al.}, 2002]{vogel2002efficacy}
Vogel, C., Cobleigh, M., Tripathy, D., Gutheil, J., Harris, L., Fehrenbacher,
  L., Slamon, D., Murphy, M., Novotny, W., Burchmore, M.  {\em et~al.}
  (2002{\em{}}) {Efficacy and safety of trastuzumab as a single agent in
  first-line treatment of HER2-overexpressing metastatic breast cancer}.
\newblock {\em Journal of Clinical Oncology, } {\bf 20} (3), 719.

\bibitem[Wang {\em et~al.}, 2005]{wang2005mcg}
Wang, P., Kim, Y., Pollack, J., Narasimhan, B.  \& Tibshirani, R. (2005{\em{}})
  {A method for calling gains and losses in array CGH data}.
\newblock {\em Biostatistics, } {\bf 6} (1), 45--58.

\bibitem[West, 1984]{west1984outlier}
West, M. (1984{\em{}}) {Outlier models and prior distributions in Bayesian
  linear regression}.
\newblock {\em Journal of the Royal Statistical Society. Series B
  (Methodological), } {\bf 46} (3), 431--439.

\bibitem[Willenbrock \& Fridlyand, 2005]{wk2005}
Willenbrock, H. \& Fridlyand, J. (2005{\em{}}) {A comparison study: applying
  segmentation to array CGH data for downstream analyses}.
\newblock {\em Bioinformatics, } {\bf 21} (22), 4084--4091.

\end{thebibliography}

\newpage
\section*{Supplementary Material}

\algsetup{
linenosize=\small,
linenodelimiter=.
}
\begin{algorithm}
\caption{Backward selection procedure for the breakpoints}
\label{alg1}
\begin{algorithmic}[1]
\REQUIRE $M_{m \times n}$, $q_\alpha$
\STATE $\mathcal{J}  \leftarrow \emptyset $ and flag $\leftarrow$ true
\REPEAT
\FOR{$j=1$ to $n$ and $j \notin \mathcal{J}$}
\STATE $V_{-j} \leftarrow m$ samples without replacement from columns $\mathcal{I}$ of $M$, $\mathcal{I}=\{i:i \neq j \mbox{ and } i \notin \mathcal{J}\}$
\STATE $V_{j} \leftarrow $ column $j$ of $M$
\STATE $\tilde{P}_j=\frac{1}{m^2} \sum_{k=1}^{m} \sum_{k^\prime=1}^{m}I(|V_{j}[k]| > |V_{-j}[k^\prime]| )$ \label{line:6}
\ENDFOR

\IF{$\exists j \in \{1,2,\dots,n \} \mbox{ and } j \notin \mathcal{J}: \tilde{P}_j > q_\alpha$}
\STATE $j \leftarrow j: \tilde{P}_j > \tilde{P}_{j^\prime} \mbox{ all } j^\prime \neq j$ \label{line:9}
\STATE $\mathcal{J} \leftarrow \mathcal{J} \cup \{j\}$ 
\ELSE
\STATE flag $\leftarrow$ false
\ENDIF

\UNTIL{flag = false  or  number of elements in $\mathcal{J}=n-1$}

\IF{number of elements in $\mathcal{J}=n-1$}
\STATE $\mathcal{J} \leftarrow \{1,2,\dots,n \}$
\ENDIF

\ENSURE $\mathcal{J}$
\end{algorithmic}
\end{algorithm}
\end{document}